\documentclass[letterpaper,english]{article}
\usepackage[T1]{fontenc}
\usepackage[latin1]{inputenc}
\usepackage{amsmath}
\usepackage{amssymb}

\makeatletter

\usepackage{graphicx}
\usepackage{amssymb}

\usepackage{amssymb}

\usepackage{amssymb}

\usepackage{graphicx}
\usepackage{geometry}

\usepackage{babel}

\usepackage{babel}

\usepackage{babel}

\usepackage{babel}
\makeatother

\begin{document}

\title{\textbf{Type Ia Supernovae and the discovery of the Cosmic Acceleration}}

\author{Alejandro Clocchiatti \\
Departamento de Astronom\'{\i}a y Astrof\'{\i}sica \\
Facultad de F\'{\i}sica\\
Pontificia Universidad Cat\'olica de Chile\\
$\&$\\
High Z Supernova Search Team }

\date{October 11, 2011}

\maketitle

\begin{abstract}
I present a review of the research and analysis paths that converged to make Type Ia SNe
the most mature cosmological distance estimator of the present time.
The narrative starts with the first works in the early decades of the 20th century and
finishes with the more recent results, covering
the surprising discovery of Cosmic Acceleration in 1998.

The review was written thinking of physicists with a strong interest in Cosmology,
who might have pondered why was that, after decades of not being able to agree upon
the rate of cosmic expansion, astronomers were so quick to concur on cosmic acceleration.

\end{abstract}


\section{Introduction} \label{intro}
Supernovae (SNe) have been with us from the beginning, since before we were ``us''.
Occasionally superimposed on the background of a familiar, regular and repetitive sky, they were among the variable phenomena that puzzled, marveled and scared our human ancestors.
The modern view associates them with the final stages of stellar evolution in scenarios where the shifting balance between pressure sources and gravitation reaches unstable regimes.
When the instabilities lead to an explosion powerful enough to disrupt the star we have SNe.

The current paradigms purport SNe as critical ingredients in the micro-physics of Cosmology.
They create the chemical history of the Universe, stir the interstellar medium of galaxies enriching primordial matter with novel chemical elements, and, if exploding in dense environments, generate shock waves that promote the birth of newer generations of stars.
In addition, from the practical point of view of today$'$s astronomers, they provide outstanding estimators for cosmologically relevant distances.

This paper is a review of the research paths that converged to make Type Ia SNe,
at the time, the most mature distance estimator for use in Cosmology, the surprising
discovery of Cosmic Acceleration in 1998, and touches the current work trying to use SNe to
obtain even finer cosmological inferences.
Opposite to what the organizers of the meeting expected, the paper was written not before, but {\em after} the presentation at the workshop.
The positive side of having done it this way, is that I could tune it to answer the questions I received after the talk.
It was the first time I presented the subject to an audience made up, primarily, of physicists, and the questions were different to what I was typically
used to answer.
Hence, I prepared this work thinking mainly of colleagues who are physicist that work in Cosmology.
I will touch some topics considered ``basic'' in SNe studies, which are available in other reviews or papers, but
are scattered around mainly in astronomical journals.
My hope is to present basic concepts, give an overview of the advances as they proceeded in time, and organize some relevant bibliographic sources to
facilitate further study from scholars who are not specialist in the field.

I must warn the reader that the text, as a review of SNe, is not complete.
It cannot be.
It is biased towards the developments that lead Type Ia SNe to become the outstanding
distance estimators that they are today.
The rest of the SNe appear and are mentioned just as the background out of which Type Ia SNe
appeared and were recognized.

I shall also warn the reader that I have made an attempt
to merge in this manuscript the different
styles of referencing literature used by physicists, like in Phillips \cite{p1993},
and astrophysicists, like in Phillips (1993). 
Being myself an astrophysicist I have come to appreciate the importance of having the
date of the papers inserted in the text, to figure out the pace of developments
and/or possible historical relation between different pieces of work.
The result, Phillips (1993 \cite{p1993}) is slightly cumbersome, but I hope that will suite
the prejudices and slants of both communities.

The chart of this paper is as follows.
Section 2 is a historical narrative.
In it I describe the evolution of the area that I loosely call ``SN Studies,'' from the
times of the pioneers, in the early decades of the 20th Century until all the major
advances in instrumentation, observation, analysis, and theoretical interpretation
had taken place, leaving the field poised for a major breakthrough by 1993.
In Section 3, I describe the fast developments between 1994 and 1998, when the Cosmic Acceleration was discovered.
In Section 4, I present the results of 1998,
using as an illustration the complete set of Type Ia SNe
published by the High Z SN Search Team, augmented by selected later discoveries.
In Section 5, I describe what could be called the ``Second Generation'' SN surveys
started soon after the turn of the century, with the goal of constraining the nature of
the now called ``Dark Energy''.

\section{Supernova Studies} 

\subsection{The pioneers}
The serious study of SNe started with two very influential papers by astronomers from the Carnegie Institution and California Institute of Technology. Baade and Zwicky (1934a \cite{bz1934a} $\&$ 1934b \cite{bz1934b}), proved clearly that the nova stars known from antiquity were composed by two very different classes.
One of them, which they named for the first time super-novae, was much more powerful
than the other.
They showed that the energy involved in these outbursts was equivalent to a considerable fraction of the rest mass of a star and they proposed that the neutronization of matter could be the source of energy for the process.
They issued the suggestion as a general possibility based on energy considerations as they did not present any specific scenario where neutronization could occur.
Specifically they did not focus on the, now preferred, gravitational collapse in the core of massive stars.
They also suggested that SNe could be one of the sources of Cosmic Rays, a possibility that has been explored and confirmed since.

Having associated supernovae with the final stages of the still largely unknown stellar evolution it was natural to device a SN search program.
If a survey is done, imaging nearby galaxies with hundreds to thousands of millions of stars each, giving enough time, chances are that a few of those stars will be caught in the act of becoming SNe.
Once they are discovered, a systematic program for studying the outbursts could be done just by following the evolution in time of the emitted flux and spectrum.
Using the recently commissioned wide angle 18-inch Schmidt telescope at Palomar observatory, Zwicky, together with J. Johnson, started the first systematic search for SNe in 1936.
The telescope would be used to image hundreds of galaxies in multiple epochs during the observing season and compare the new images with the old ones using blinking microscopes.
Any new star could be easily spotted.
Zwicky teamed up with another astronomer of the Carnegie Institution to follow up the spectroscopic evolution.
After a few years of work, Minkowski (1941 \cite{m1941}) was able to start the dissection of the SN events classifying them into ``types'' according with their spectra, a tendency that has only partially helped to clarify the field, but, in any case, has been impossible to change.

Minkowski proposed that SNe come in two different spectroscopic types, I and II.
The early time spectra of Type I SNe were not understood.
They showed wide absorption/emission bands that shifted in position and strength with the passage of time.
But it was not possible to identify the chemical elements that caused them.
It was clear, though, that they were a homogeneous class because the unknown spectra would repeat from one event to the next.
A few spectra taken hundreds of days after outburst provided broad features that could be reasonably associated with emission lines,
out of which only one forbidden transition of O~I was identified.
The evolution in time was also remarkable similar for different SNe within this type.
Type II SNe, on the other hand, displayed the well known lines of the hydrogen Balmer series and the He I line at 5678 angstroms.
It was easy to recognize the characteristic shape of P Cygni profiles and, hence, measure expansion velocities.
These turned out to be in the order of several thousand kilometers per second.
Thus, the hypothesis of an explosion with massive ejection of matter was confirmed for Type II SNe and,
by extension, adopted as a reasonably working hypothesis for Type I as well.

Zwicky also contributed to the initial proliferation of SN types by using the characteristics
of the photographic light curves (the light curve is the evolution of brightness in a given
photometric band with time) as a tool to aggregate SNe into classes.
This idea also stuck, although the groups that he proposed have been abandoned (see for example a later recollection in Zwicky, 1965 \cite{z1965}).

Classification of events aside, the photometric follow up provided a puzzling result.
The SNe sustain the emission of energy for hundreds of days.
Type I, in particular, after they go through the maximum emission of light during the initial weeks, settle on a linear decay of brightness with time
if plotted as magnitude (i.e. logarithm of the emitted energy).
This exponential decay phase was followed for hundreds of days in some events.

As the database of observations increased, the spectra of Type I SNe continued to be a source of perplexity.
It was not even known how to interpret them in the most basic terms.
Were they emission line spectra of atoms in different ionization states whose lines were enormously broadened by large expansion velocities?
Were they combined spectra, where a continuum and broadened absorption and emission lines are present,
like an extreme case of some peculiar stars or normal novae?
The first serious attempt to make sense of Type Ia spectra was by Payne-Gaposchkin and Whipple (1940 \cite{pw1940}) and
Whipple and Payne-Gaposchkin (1941 \cite{wp1941}).
They attempted to reconstruct the SN spectra by summation of the emission lines of the more common chemical elements in astrophysics, allowing for several stages of ionization.
They produced synthetic spectra that were a very rough match to the observations and left open the possibility that many of the troughs in the spectra were real absorption minima that they had not tried to model.

The pioneering work described above created the field of SN studies and set the landscape and main inroads for its subsequent evolution.
The early discoveries were puzzling, and left broad and important questions to be answered.
The neutronization of matter was an fascinating possibility for the production of the energy required to disrupt the stars,
but there were no scenarios identified for such a process to occur.
In addition, the fact that there were two SN types suggested that there could be two different explosion mechanisms.
The early attempts at theoretical interpretation of SNe spectra explored for the first time what, in the end, would be the key approach to decipher them: the construction of synthetic spectra.

\subsection{Getting the picture: The 40s and 50s}
The 40s and 50s were a period of slow accumulation of observations and evolution of the conceptual tools to understand, in general, astrophysics.
Basic knowledge on atoms, atomic nuclei, nuclear reactions, and improvements in the measurements of the cosmic abundances, helped to establish the cosmological role of stars and stellar nucleosynthesis.

The data of Zwicky$'$s first survey continued to be studied, until they started to reveal its full content.
An important event, which will play its role later on, was the discovery and observation of a bright SN in NGC 4214. This SN was observed from Europe and the US and Wellmann (1955 \cite{w1955}) remarked that it was of Type I, but the spectrum was somehow peculiar (we know now that it was the first Type Ib SN with spectra recorded).

Another relevant piece of observational information that finally came to be recognized was the outstanding light curve of SN 1938C observed by Zwicky, and presented by Baade (1945 \cite{b1945}).
The part of the light curve that comes after maximum, the exponential decay, had a half-life of $\sim$55 days.
This slope was essentially the same for all Type I SNe and became a strong argument in favor of a common energy source for all Type Ia SN.
For SN 1938C, the brightest Type Ia SN since the invention of the telescope and the better
observed event at the time,
the exponential decay lasted as long as the observations, more than 10 times the half-life.
Borst (1950 \cite{b1950}) squarely set the focus on this fact and concluded that it was the natural result a radioactive decay.
Connecting directly the half-life of the SN light curve with the half-life of the unstable nuclei, he suggested that $^7$Be was the energy source and proposed a mechanism to form this unstable nucleus following the collapse of the stellar core after hydrogen exhaustion.
Baade et al. (1956 \cite{bbhf1956}) and Burbidge et al (1956 \cite{bbf1956}) picked up the idea, emphasizing that in addition to $^7$Be, both $^{89}$Sr,
and the recently discovered $^{254}$Cf, were also good matches for the half-life time, and they justified their preference for the latter as the
energy source for SNe.

Observations of abundance of elements in meteorites, Earth, the Sun, stars and nebulae, also started to converge and it was possible to interpret them in terms of the recently created nuclear shell model.
Suess and Urey (1956 \cite{sandu1956}) published a paper that set the standard of the time, with the first modern looking plots of the relative abundances of chemical elements.

Most of the ideas and advances of the last decades that were slowly converging to the big picture of astrophysics were organized in one of the more influential papers of the 20th century astrophysics by Burbidge, Burbidge, Fowler and Hoyle (1956 \cite{bbf1956}, hereinafter named B2FH).
In this paper, they presented a coherent picture for the origin of the chemical elements in thermonuclear reactions of increasingly heavier nuclei in the core of evolving stars.
They also emphasized the need for additional neutronization of complex nuclei,
and developed the concept of slow and rapid neutron capture processes.
The slow processes would take place sometime during normal stellar nucleosynthesis and the rapid ones in the explosions of SNe.
It was, essentially, the modern view, with a few notable shortcomings that are easy to pinpoint now, 55 years after.
Many of those were related with stellar astrophysics and SNe.
In the effort to provide the synthesis authors disregarded the fact that the two spectroscopic SN types could point out to the existence of two different progenitors and/or explosion mechanism.
B2FH correctly located the terminal instability of core-collapse SNe in the $^{56}$Fe core of massive stars that had followed all the nucleosynthesis sequence, but did not have a clear understanding of the processes that triggers the collapse.
They did not have, either, a clear understanding of how the core collapse would result in the ejection of the rest of the star and the explosive nucleosynthesis.
Their concept was that the core collapse will start a rapid compression inwards of the whole star, including the mantle.
This would sharply raise the temperature of gas of the mantle, rich in nuclear fuels,
and trigger a runaway thermonuclear reaction which will disrupt the star.
B2FH also insisted on $^{254}$Cf as the radioactive product that would power the light curves, at times as if the presence of $^{254}$Cf in SNe were an observational fact.

\subsection{Getting the picture right: The 60s and 70s}
The 60s started well for SN science.
Hoyle and Fowler (1960 \cite{hf1960}), recognized the importance of light nuclei  ($^{12}$C, $^{16}$O, $^{24}$Mg, etc.) as thermonuclear fuel when they were in highly degenerate conditions.
They identified degenerate low mass stars near the Chandrasekhar limit as an optimal scenario for explosions sustained by these fuels, and associated these explosions with Type I SNe.
They estimated the mass of terminally unstable stars and found it to be constrained to a very narrow range.
Out of this fact they explained the relatively scarcity of the explosions, but failed to recognize the, even more remarkable, relevance for the homogeneity of the class.
This is the first paper where collapse of evolved cores and thermonuclear explosions are clearly associated with Type II and Type I SNe, respectively.
Hoyle and Fowler revisit as well the $^{254}$Cf hypothesis and found it not as compelling as in previous papers, because other radioactive heavy nuclei had been discovered with similar half-lives.
Also, for the first time clearly, they stressed that the half-life of the powering nucleus need not be directly matched by the half-life of the SN light curve.

Colgate (see for example Colgate et al. 1961 \cite{c1961}), Arnett (1966 \cite{a1966}), and Truran (1966 \cite{t1966}) started a long series of quantitative studies of spherical shocks running in stellar envelopes, the associated nucleosynthesis, and how their energy output could match the observed SN mass ejection and light curves.
An important advance came after Colgate and White (1966) realized that the B2FH paradigm of thermonuclear explosion induced by collapse of the core in massive stars does not work, because the collapse actually starts a rarefaction wave at the inner part of the mantle that quenches thermonuclear reactions.
What they found, in turn, was that the dynamical implosion of the evolved core
was so violent that a vast amount of energy, many times greater than the thermonuclear one, was available just from the deepening of the
gravitational potential well.
This energy was transferred to the mantle via the emission and deposition of neutrinos.
This was the birth of the ``prompt shock'' model for core-collapse SN explosions.

Truran, Arnett and Cameron (1967 \cite{tac1967}) studied the process of Si burning.
Afterward, Truran himself suggested Colgate to consider the consequences of Si burning in his model light curve calculations.
Colgate and McKee (1969 \cite{cm1969}) presented the first numerical results where theoretical light curves are a reasonable match to observations.
The importance of the $\alpha$-particle isotope $^{56}$Ni, which decays into $^{56}$Co with a half-life of 6.1 d, and then into $^{56}$Fe, with a half-life of 77.12 d, had been found.
This was conceptually very important to make astronomers realize that essentially the full display of Type I SNe was due to radioactive decay, and not just the exponential tail.

Finzi and Wolf (1967 \cite{fw1967}), picked up the thermonuclear runaway in degenerate matter proposed by Hoyle and Fowler (1960 \cite{hf1960}).
They focused on massive white dwarfs as likely candidates for a catastrophic change that would trigger the explosion,
thought of those SNe that have been observed in elliptical galaxies, where star formation ended more than $10^{10}$ years ago,
and asked the right question:
How can a star that reached the white dwarf stage so long ago suddenly become a Type I SN?
They went on trying to answer it assuming that very slow electron captures dynamically change the dwarf
star and trigger collapse, heating, and thermonuclear runaway.
For the process to work, they had to build white dwarfs of fairly exotic composition.
Hansen and Wheeler (1969 \cite{hw1969}) took the idea a step further and numerically computed the explosion of a white dwarf of $^{12}$C.
They showed that the collapse and nuclear detonation was sufficiently catastrophic so as to explode the whole star providing energies and ejection velocities that matched the observations.
Wheeler and Hansen (1971 \cite{wh1971}) expanded the calculation to $^{12}$C and/or $^{16}$O and computed the nucleosynthesis products.
They found that the whole nuclear fuels were burnt all the way to nuclear statistics equilibrium (i.e. all iron group elements).
They also proposed the model of a white dwarf close to the Chandrasekhar mass accreting matter from a close binary companion as a possible scenario for Type I SNe.
Truran and Cameron (1971 \cite{tc1971}) seem to have reached the idea independently, and they went on to postulate that $^4$He ignition in the accreted
matter will trigger the $^{12}$C runaway inwards, and then the explosion (this is still considered a possible explosion mechanism for some Type Ia SNe).
Arnett (1969 \cite{a1969}) also explored numerically the detonation of a $^{12}$C white dwarf, or white dwarf-like core in an intermediate mass star.
A well as Hansen and Wheeler in the previous papers, he found that it created too much iron group elements, a fact that was probably inconsistent with the chemical history of the Galaxy.
He concluded that not many SNe could be of this type.
Arnett realized, however, that the nucleosynthesis products were very sensitive to the critical density at which $^{12}$C ignites, and stressed that a lower value of this density would help to produce elements between $^{12}$C and $^{56}$Ni, instead of just iron peak ones.
Nomoto, Sugimoto and Neo (1976) asked the critical question: Who knows that the detonation is actually initiated [in a C-O white dwarf]? They proposed that the thermonuclear combustion could take the form of a subsonic deflagration, develop convective instabilities, evolve violently outwards and anyway disrupt the star, without ever entering the detonation regime.
They found that, if the deflagration is slow, the combustion proceeds in two phases, the second of which starts when the star has expanded.
This second phase takes place at a lower density and avoids overproduction of the iron peak elements.

Regarding the understanding of SN spectra, McLaughlin (1963 \cite{md1963}), an expert on Novae, revisited the spectroscopic plates
of SN~1954A that had been taken at Lick Observatory and mostly neglected afterward.
He was interested by the remark from Wellmann (1955 \cite{w1955}) that the SN was ``peculiar'', and decided to make his own line identifications.
After carefully looking at all the recorded spectra he concluded that they were combined emission and absorption spectra.
After coming up with the concept of absorption-like and emission-like features, he thought that some spectra looked similar to those of B stars, with little or no hydrogen.
He recognized that a couple of absorption-like minima in the blue region, if identified with
He I lines, gave consistent velocities, and went on to identify many other features all
shifted by velocities of $\sim$5000 km s$^{-1}$,
and broadened by velocities of order 10$^3$ km s$^{-1}$.
Minkowski (1963 \cite{mr1963}), focusing mostly in the identification of emission-like features, criticized the approach.
Pskovskii (1969 \cite{py1969}) recognized the value of both the realization of the dual (absorption and emission) character of the spectrum by McLaughlin, and the stress of Minkowski on how uncertain was the identification of emission-like features.
He realized that the important issue was that, finally, {\em absorption lines had actually
been identified upon a continuum},
and applied the idea to the still unknown spectra of the normal Type I SNe.
He struck gold.
His paper reveals for the first time that Type I SNe display low excitation lines of low ionized species, with no trace of light elements.
He first correctly identified the, now, characteristic lines of Si II, and then went on to find Fe II, Mg II, Ca II, and S II.
This pattern of lines has hold since then.
His, more doubtful, identification of He I, has been superseded.
Mustel (1971a \cite{m1971a}, 1971b \cite{m1971b}, 1972 \cite{m1972}), provided some complementary identifications, and some differing interpretation.
Branch and Patchett (1973 \cite{bp1973}) settled the issue in a very elegant way, developing the first modern synthetic spectra and comparing them
with the better observed ones.

It was in this period, as well, that SNe were first identified as valuable light sources for cosmological estimation of luminosity distances.
Kowal (1968 \cite{k1968}) collected the first sample of 22 well observed Type I SNe with light curves in photographic magnitudes, found that they had a scatter of $\sim$0.6 mag at
maximum, and proposed a program to find SNe in clusters of galaxies with the goal of
reducing the uncertainty up to 0.1 or 0.2 magnitudes by averaging many of them.
He pointed out an important fact.
Type Ia SNe were powerful sources of light but of stellar origin.
As such, they were very probable not subject to the strong evolutionary effects that systematically biases the luminosity distances computed using objects that are aggregates of stars.
Both galaxies and clusters of galaxies are evolving very fast with time in the region of the Universe available for observation.
But stars have been fairly similar to themselves for thousands of millions of years.
He was visionary enough to suggest that when light curves for very distant Type I SNe
became available, they could be used to know not only H$_0$, the Hubble constant,
but also ``the second-order term in the redshift-magnitude relation''
(i.e. the deceleration parameter).

By the end of the 70s, the scenarios to produce Type I SN progenitors and explosions were fairly sophisticated.
The match between theory of spectra and observation started to make sense and provided a tension which would not be solved for a few years: Theoretical explosions produced mainly iron group elements,
but the observed spectra displayed $^{56}$Fe but also appreciable amounts of intermediate
mass elements, which were not produced by the theoretical models.
On the other hand, Type I SNe had been identified as remarkable candidates for measuring cosmologically relevant distances, providing a strong incentive to programs for their search and study.
By the end of the decade both Colgate (1979 \cite{c1979}) and Tammann (1979 \cite{t1979}) proposed cosmology with Type I SN as one of the major science drivers for the future Space Telescope.

\subsection{Getting the picture in focus: 1980-1993}
The early eighties witnessed very fast developments in SN science.
Some were triggered by observations.
In 1981 a SN was discovered in NGC 4536, it was a Type I event that was followed from McDonald Observatory with good wavelength and time sampling up to more than a hundred days after maximum light.
Large telescopes and modern electronic detectors were used, which resulted in the best SN spectra ever obtained.
The data was analyzed by Branch et al. (1983 \cite{blm1983}) who produced a paper that set the standard for theorists to compare with.
Both spectroscopic observations and synthetic spectra confirmed the picture that the $^{56}$Ni $\rightarrow  ^{56}$Co$\rightarrow  ^{56}$Fe was required, and that, in addition to the radioactive Ni, a sizable fraction of Si, Ca, Na, S, Mg and O were also ejected.

The improvements in detector technology continued to play a role.
Elias and collaborators started a program to follow up SNe in the infrared from Cerro Tololo Inter-American Observatory (hereinafter CTIO; Elias et al. 1981 \cite{efh1981}).
By 1985 they had found that the infrared eye saw two different types of light curves among SNe of the type I, where the optical view had seen just one.
Elias et al (1985 \cite{emn1985}) coined the names ``Type Ia'' for the more frequent, bluer, and typically brighter SNe and ``Type Ib'' for the other.
Spectroscopy at McDonald Observatory indicated that the two classes had different spectra (Wheeler and Levreault, 1985 \cite{wl1985}).
It took few years to identify strong He I lines in the spectrum of Type Ib SNe (Harkness et al. 1987 \cite{h1987}),
and a few more to realize that there were some peculiar Ia spectra with neither Si~II nor He~I lines,
making it natural to create the Type Ic bin for them.
The advances in the field by the end of the decade are reviewed by Wheeler and Harkness (1990 \cite{wh1990}).

The eighties also saw the start of the first SN surveys based on automatic telescopes with digital detectors.
Some of them targeted local SNe and some of them were specifically designed to crop distant SN for using as cosmological proves.
Some, in addition, planned an automatic pipeline of data reduction and image comparison to detect SNe candidates with minimal human intervention.
One of these first initiatives was the Berkeley Automated SN Search (BASS, Kare et al. 1981 \cite{kpm1981}).
The BASS took many years to fly.
Perlmutter (1989 \cite{p1989}) reports the status after (almost) the first year of real time automatic operation.
At about the same time the BASS also started to plan and design a search for more distant SN to use in cosmology (Couch et al. 1991 \cite{cpn1991}).

The first group to take the leap, and really go after distant SNe to set constraints on cosmology was a Danish group (Hansen et al. 1987 \cite{hjn1987}).
The search was based on the 1.5m Danish telescope at ESO La Silla, in Chile.
It was already a modern search with CCD cyclic imaging, and real time data analysis including image matching and subtraction.
Due, mainly, to the small size of the detector, the impossibility of using the $R$
photometric passband band due to strong fringing (this caused by standard interference
in the thin silicon layers of the detectors, and was usual in the older CCDs but
greatly improved along the nineties), they discovered only one useful SN in two years
and then dropped the program (Noorgard-Nielsen et al. 1989 \cite{nhj1989}).

Looking back at their effort is clear that they were way ahead of the proper time.
But the finding of SN 1988U at z=0.31 proved that the strategy of multi epoch imaging and digital image comparison did lead to discover distant SNe.
By the early nineties the Berkeley SN team made an agreement to put a large CCD camera on the 2.5m Isaac Newton Telescope in the Canary Islands in exchange for observing time to search for distant SNe.
It was this combination that allowed them to discover SN 1992bi at z=0.46, the more distant SN up to that time (Perlmutter et al. 1995 \cite{p1995}).
After that success the group started to compete for time at the run of the mill 4.0m class telescopes of the world, and finding distant SN became more usual.

Another critical development of the time was the spread of large CCD detectors around telescopes at different observatories.
This made it possible to follow up SN even in non photometric conditions.
Since it was typical that the parent galaxy and the SN would appear in a field together with some Galactic foreground stars, relative photometry could be done with respect to a local standard sequence.
By the early nineties, good quality, well sampled, multicolor light curves of SNe in nearby galaxies started to accumulate.
A compilation by Sandage and Tamman (1993 \cite{st1993}) found that the scattering of the B maximum brightness of SN Ia, treated as if they were standard candles, around the linear relation implied by the Hubble flow was 0.51 mag.
This was a modest improvement from the 0.6 mag found by Kowal in 1968, but typical of using heterogeneous samples of Type Ia SN as plain standard candles.

At about the same time, Mark Phillips, at CTIO was working in a different direction.
Pskovskii (1977 \cite{py1977}, 1984 \cite{py1984}) had proposed to use the rate of decline after maximum as an additional element to classify SNe, of all types, and suggested that there were correlations between the decline rate and other SN properties.
Phillips (1993 \cite{p1993}) attempted the same kind of study but now with modern data and a much cleaner sample of real Type Ia SNe.
He found a strong correlation between the rate of decay after maximum light and both the absolute magnitude of SN Ia at maximum, and the $B-V$ color (an index of the SN temperature).
The correlation was strongest in the $B$ band, moderate in $V$, and minor in $R$.

This was a very important discovery: Type Ia SN were only approximate standard candles.
Yes, there was a sizable intrinsic scattering of the luminosity at maximum light, but it was possible to recognize how far away from the mean value an individual SN was, just from the rate of decay after maximum.
As important as this, the same rate of decline was an indicator of the intrinsic color of the SN.
This made it possible to correct for the extinction of light caused by foreground interstellar matter in the parent galaxy.
Both corrections have a direct impact in improving the character of Type Ia SNe as cosmologically relevant distance estimators.
Phillips (1993 \cite{p1993}) was a major breakthrough.

There were also some hits on the theoretical side, but the advances were limited by
computer power.
Nomoto, Thielemann $\&$ Yokoi (1984 \cite{nty1984}) following up the idea of Nomoto, Sugimoto $\&$ Neo (1976 \cite{ns1976}), computed models of carbon deflagration supernovae.
One of the main problems they faced was that the deflagration gives raise to Rayleigh-Taylor instabilities and is, hence, a multidimensional process.
Forced to treat it in one dimension, they simulated the propagation of the convective carbon deflagration front using a time dependent mixing length theory.
They had to assume several numerical parameters to accommodate the hydrodynamics, but followed the nucleosynthesis in detail.
One of the models they computed, W7, was a very good match to the observations.
W7 was a complete success in terms of explosion energetic, both in the time dependent luminosity output and kinetic energy transferred to the ejecta, but also, for the first time in producing the right amounts of intermediate mass elements at the velocities seen in the early time SN spectra.
In particular, synthetic spectra computed using the output produced by the model gave a good fit to the observations,
if the outer layers of the SN were mixed during the late stages of the explosion (Branch et al. 1985 \cite{bdn1985}, Nomoto en al. 1986 \cite{nty1986}).
The latter was needed, because the intermediate mass elements in the theoretical model appeared in a very narrow range of velocities, smaller than those observed in real SNe.
In spite of this problem, W7 set the standard of comparison for Type Ia SN models for many years to come.
Woosley and Weaver (1986 \cite{ww1986}) produced a very influential review paper on
SNe in general, centered especially on the physics of explosion and nucleosynthesis,
which remains a relevant reference even today.

In addition to the location of the intermediate mass elements, the deflagration models
had other important shortcoming: They tended to overproduce $^{54}$Fe and other neutron
rich Fe peak isotopes, because the burning matter remains at very high temperatures and densities for too long.
Also, upon further study, the need to mix the outer layers to take intermediate mass elements to higher velocities became increasingly more difficult to justify (Sutherland and Wheeler 1984).
A plain deflagration explosion model, in addition, seemed too constrained
to explain the heterogeneity of Type Ia SNe that was starting to appear with improved observations.
Khokhlov (1991a \cite{k1991a}, 1991b \cite{k1991b}) focused on these problems, and decided to test whether reality could be more complex.
He realized that the subsonic nature and related time scales of the deflagration were long enough to change the background where thermonuclear burning was taking place, and, hence, a detonation was likely after a period of deflagration.
He named the mechanism delayed detonation and showed that a white dwarfs exploding like that could match the observations better than the plain deflagration of W7.
Later study proved that the combination of processes had another virtue: varying the critical density at which the deflagration turned into a
detonation (an unknown external parameter in Khokhlov approach), different kinds of Type Ia SNe could be produced matching the diversity of
normal, sub-luminous, and luminous events, giving rise to a theoretical interpretation of the Phillips (1993) relation
(H\"{o}fflich, Khokhlov $\&$ Wheeler, 1993 \cite{hkw1993}).

Looking with hindsight, by the end of 1993 the field was poised for a breakthrough.
Type I SNe had been cleaned so as to clearly isolate the Type Ia events.
Theory had advanced enough to solidify the hypothesis that Ia SNe originate on the thermonuclear combustion of a Chandrasekhar mass white dwarf.
This gave strong support to the concept that Ia SNe ought to be very uniform.
Reality showed that, nevertheless, there were sources of heterogeneity, but empirical study prompted to a feasible path at calibrating the differences.
In addition, theory parametrization of the unknown critical density for a deflagration
to turn into a detonation, allowed for a qualitative understanding of these differences,
bringing some peace of mind to observational cosmologists who were using the empirical
relations without fully grasping its meaning.
Solid state detectors were large enough to allow astronomers do relative photometry between extragalactic SN and foreground galactic stars in the field.
This made it possible to do extensive follow up even in non photometric conditions.

Finally, the incipient and fast improving
internet was a development, not fully recognized at the time, which resulted critical.
It allowed astronomers at different observatories and research centers in the world to
share results, images and information in real time, making it possible to pool
human resources, and observational and computing facilities.
This was possible with modest budgets and widespread and heterogeneous funding sources.
Hence, advancing a large program of observation and analysis of distant SNe was no longer
restricted to large and rich research groups or labs, which could count with
all the facilities needed under a centralized management, but became within reach
of more horizontal arrays of researchers who could coordinate facilities scattered
around the world to work on the same project.

\section{A magic SN Cosmology lustrum: 1994-1998}
Some groups of SN studies
recognized the breakthrough implied by the Phillips (1993 \cite{p1993}) result.
Important as it was, it was based on a small and heterogeneous sample of SN observed with different telescopes and instrumental sets.
It was critical then to increase the SN sample and, ideally, observe them under more uniform conditions.
Some surveys were started with the specific goal of calibrating the relations of rate of decline versus magnitude, and color, at maximum light.
Among the first were the Cal\'an/Tololo Survey (Hamuy et al. 1993 \cite{h1993}), where M. Phillips himself was a key player, and the long
standing SN study program by B. Kirshner, students and post-docs at the Harvard-Smithsonian Center for Astrophysics (CFA).
Also, several methods to accomplish the calibration were developed.
Among them, the $\Delta m_{15}$ (Phillips 1993 \cite{p1993}, Hamuy et al. 1996 \cite{hetal1996}, Phillips et al. 1999 \cite{ph1999}), and the Multicolor Light Curve Shape method, by Riess, Press and Kirshner (1995 \cite{rpk1995}, 1996 \cite{rpk1996}) at Harvard-CFA.

As mentioned earlier, the SCP had started to be effective at finding distant SNe.
By 1994 they reported six SNe discovered in approximately six nights of a program combining the 2.5m Isaac Newton and the 4.0m KPNO telescopes (Perlmutter et al. 1994 \cite{p1994}).
In 1995 (Goobar $\&$ Perlmutter, 1995 \cite{g1995}), the group presented a detailed account of an elegant method to use the luminosity distances measured towards SNe to constrain the cosmological parameters $\Omega_M$ and $\Omega_\Lambda$.
The method builds contour probability levels comparing cosmological model predictions with an observed Hubble diagram of nearby and distant Type Ia SNe.
The paper became a model for the methodology, and most subsequent analysis built up from this study.
It showed, however, that the group was still not giving serious quantitative consideration to the ``Phillips effect,'' and to the reddening correction of individual SNe.

It is my personal view, that it was in part the perception of these weaknesses
in the analysis planned by the SCP that motivated,
by the end of 1994, Brian Schmidt, a postdoctoral fellow at the at Harvard-Smithsonian
CFA, and Nick Suntzeff, a senior staff astronomer at Cerro Tololo Inter-American Observatory,
to launch an independent enterprise to build a sizable sample of distant SNe.
The group was called ``High Z SN Search Team'', to emphasize a horizontal constituency.
The main goal of this new group was to observe the SNe in a way that allowed a sine qua non spectroscopic classification, a careful understanding of the interstellar extinction in front of each individual SN, a precise measurement of the light curve shape, and a minimization of the systematic effects associated with $K$-corrections.
The first telescope runs of this team were in March 1995, when they discovered SN 1995K at z=0.478, the record holder of the time (Phillips et al. 1995 \cite{pd1995}).
I joined the group in September 1995, when moved to CTIO as a postdoctoral fellow.

The years 1996-1997 were thrilling.
The SCP and the HZ Team routinely presented proposals to discover and follow up SNe from different telescopes around the globe, including the Hubble Space Telescope.
The SCP managed to build what they thought was a small but trustable sample by mid 1997, when they published the paper  Measurement of the Cosmological Parameters Omega and Lambda from the First Seven Supernovae at z~$\gtrsim$~0.35 (Perlmutter et al. 1997 \cite{p1997}).
In this paper they present the elegant Stretch Factor Parametrization, and independent way of calibrating the light curve width versus luminosity effect found by Phillips in 1993.
They, however, could not correct for extinction the individual events and some of the SN did not have a clear spectroscopic classification as ``Type Ia''.
The results pointed towards a high density universe, where a cosmological constant was excluded with high statistical significance ($\Omega_\Lambda <$ 0.51 at the 95\% confidence level for a spatially flat universe).

By 1998 the results of the HZ Team started to appear (Garnavich et al. 1998a \cite{g1998a}; Riess et al. 1998 \cite{retal1998}; Schmidt et al. 1998 \cite{schmidt1998}; Garnavich et al. 1998b \cite{g1998b}) and the times went from thrilling to hectic.
The first paper was based on a set of four SNe, three of them observed with HST, and was also presented at the January, 1998, meeting of the American Astronomical Society (Garnavich et al.  1997 \cite{g1997}).
It arrived squarely at the opposite conclusion that Perlmutter et al. (1997 \cite{p1997}) did: The SN distances were more consistent with a low density Universe that will expand forever.
Also by the end of 1997, the SCP was starting to change its view regarding the high density Universe.
A single SN well observed with HST and added to the sample of seven SNe published earlier had changed the confidence levels enough to turn their claim into ``results [that] are preliminary evidence for a relatively low-mass-density universe'' (Perlmutter et al. 1998 \cite{p1998}).
Schmidt et al. (1998 \cite{schmidt1998}) describes the strategy and analysis method of the HZ Team, and applies them, as a test, to SN 1995K.
Riess et al. (1998 \cite{retal1998}) brought the field to a climax presenting the analysis of the first 16 HZ Team distant SNe, with the unavoidable conclusion that the expansion of the Universe is accelerating.
A paper submitted in September 1998 by the SCP presented the results based on 42 SNe, and reached essentially the same
conclusion (Perlmutter et al. 1999 \cite{p1999}), confirming finding of an acceleration.
The final work of the year by the HZ Team (Garnavich et al. 1998b \cite{g1998b}) set the first constraints on the parameter $w$ in the equation of state of the Universe, based on SN luminosity distances.

The discovery of the Cosmic Acceleration made a big impact.
It was selected as the ``Science Breakthrough of the Year'' for 1998 by Science Magazine
(Glanz 1998 \cite{glanz1998}).
It later meant an impressive sequence of prizes for Perlmutter, Riess and Schmidt, and,
in the case of the 2007 Gruber Cosmology Prize, also an explicit acknowledgement for the teams.
The string of recognitions culminated in October 2011,
with the award of the Nobel Prize in Physics to Perlmutter, Riess and Schmidt.

\section{Cosmic Acceleration: The SN result}

\subsection{``Theory''}
Given in the context of a meeting that dealt mostly with modifications of General Relativity that could help to explain perplexing
observational results, the following paragraphs seem children's play. However, General Relativity is the framework against which the observations
of distant SNe have been compared and the parameters that came out of that comparison have been the source of our puzzlement.
So, let us become children again, go back to run-of-the-mill General Relativity and assume, in addition, that the Universe is isotropic and homogeneous.
With these hypothesis, the luminosity distance between us and an object located at redshift $z$ can be written
(Carroll, Press and Turner, 1992 \cite{cpt1992})

\begin{equation} \label{eq:D_L}
D_L = \frac{c(1+z)}{H_0 \sqrt{|\Omega_k|}} \, \, {\rm sinn}
\left\{ \sqrt{|\Omega_k|} \int_0^z
\left[ (1+z)^2 (1+\Omega_M z) - z (2+z) \Omega_\Lambda \right]^{-1/2} dz'
\right\},
\end{equation}
where $H_0$ is the Hubble constant, $\Omega_M$, $\Omega_\Lambda$, and $\Omega_k$ are the cosmological
density parameters for gravitating matter (including baryonic and dark matter), Cosmological Constant, and curvature, respectively,
measured in terms of the critical density, and

\begin{figure}
\centering
\includegraphics[height=10cm]{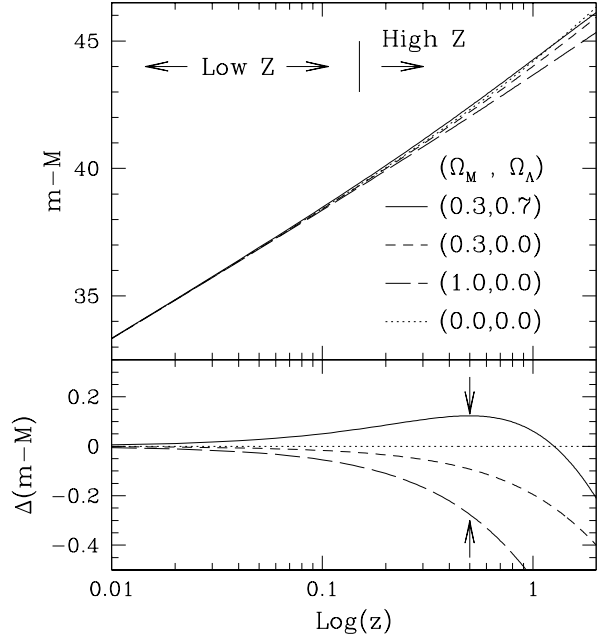}\vspace*{10mm}
%
%
\caption{Hubble Diagram of Luminosity distances for various models of the Universe, parametrized by the values of the cosmological
density parameters $\Omega_M$, $\Omega_\Lambda$, under the assumption that the curvature is zero. The upper panel displays the run of $D_L$
and the lower the difference between a given universe and the empty one.
In the upper panel the regions considered throughout this paper as low and high redshift are indicated. The vertical arrows in the lower panel
mark the position of the critical $z = 0.5$ redshift.
} \label{fi:D_L}
\end{figure}

\begin{equation}
{\rm sinn}(x) = \left\{ \begin{array}{rcl}
\sinh{x} & \mbox{for} & \Omega_k > 0 \\
x        & \mbox{for} & \Omega_k = 0 \\
\sin{x}  & \mbox{for} & \Omega_k < 0
\end{array}\right.
\end{equation}

It is standard usage in astrophysics to express the distances as {\em Distance Modulus}. This is the difference between the
apparent and  absolute magnitudes of a source, related with distance by
\begin{equation} \label{eq:dmod}
m-M = 5 \log{\frac{D_L}{10}},
\end{equation}
where $D_L$ should be given in parsecs. An advantage of the distance modulus is that it is measured directly in magnitudes, so
differences in $m-M$ can be directly compared with uncertainties in astrophysical observations and calibrations.

A rapid analysis of how $D_L$ varies with redshift helps to explain the opportunity that
the HZ Team founders perceived after the Phillips (1993 \cite{p1993}) result.
Figure \ref{fi:D_L} shows $D_L$ for some of the different values of the cosmological
parameters $\Omega$ that were
seriously considered in the early nineties, and, in addition,
the values of the now called {\em Concordance} Universe.
As became usual in the field, the upper panel displays the run of $D_L$ for different models of the universe, and the lower panel the difference
between a given universe and a reference one. Typically the reference universe is the empty, or coasting, universe.
It is easy to see in the lower panel of the figure that the difference between models at redshifts of $z \sim 0.5$ is several tens of magnitude.
So, with a dispersion of a $\sim 0.15$ mag per SN, a few tens of SNe at this critical redshift would provide an observation precise to a few hundredth
of a magnitude. This point will permit to discriminate between different cosmological models.
On the other hand, technology had improved enough by $\sim$1994 that discovering and observing Type Ia SNe at $z \sim 0.5$ was no longer the heroic
enterprise that N{\o}rgaard-Nielsen et al. (1988 \cite{nhj1989}) had undertaken. The CCDs were larger, more sensitive and, especially, did no have fringing in the
red. The latter allowed for a clean image subtraction at the rest-frame blue pass-band at $z \sim 0.5$. This allowed for precise photometry at the
pass-band where the light curve shape versus luminosity relation was most sensitive. Finally, the National Optical Astronomical Observatory IRAF group
had started to develop image matching and subtraction software by the late eighties, and by the early to mid-nineties working moduli became available as
prototype IRAF tasks (Phillips and Davies 1995 \cite{pd1995}).

On the other hand, although there is room to discriminate, it is also clear from the figure that systematic effects that change the individual SN
results by tenths of a magnitude should not be missed, especially if those effects make you prone to Malmquist bias.
Phillips (1993 \cite{p1993}), Hamuy (1994 \cite{hetal1994}, 1996 \cite{hetal1996}), and Riess, Press and Kirshner
(1995 \cite{rpk1995}) had shown that Type Ia SN considered normal by any
standard had differences in maximum brightness of that order. A magnitude limited
search for distant SNe would preferentially discover bright events at large distances
and more average SNe nearby.
The same would happen if corrections by foreground reddening, which amounts typically to few tens of a magnitude, were not considered.
Searches will tend to discover less extincted SNe at large distances and more average extincted SNe nearby.
So, if SNe were treated as standard candles and corrections by light curve shape and foreground extinction were not applied,
the final comparison would be made between populations of different intrinsic brightness
at different redshifts, and the resulting cosmology would be biased.

Finally, equation \ref{eq:D_L}, \ref{eq:dmod} and Figure \ref{fi:D_L} allow us to see the
key difference between the experiment of measuring the rate of expansion, H$_0$, and
estimating $\Omega_M$ and $\Omega_\Lambda$.
Measuring H$_0$ requires calibrating very precisely the absolute magnitude of, in this
case, Type Ia. This is, requires precise knowledge of $M$.
In turn, this calls for accurate measurement of the distance, in physical units, to many
galaxies that have hosted Type Ia SNe.
Estimating the $\Omega$ parameters, on the other hand, requires just a relative measurement.
Variations of $M$ and H$_0$, constants that could be packed together in equation
\ref{eq:dmod}, will change the vertical scale of Figure \ref{fi:D_L}, but will not change
the relative shape of the different curves.
This means that, given the appropriate combinations of telescopes, detectors and computers,
estimating H$_0$ is a more difficult experiment than estimating the $\Omega$ parameters.

\subsection{Observations: High precision, high redshift Hubble Diagrams}

By the end of 1997, the group had discovered more than 20 transients and 14 of them had been positively identified as Type Ia SNe with spectra
taken from the Keck Telescope, the Multiple Mirror Telescope, or the European Southern Observatory 3.6m.
This set, which included the four SNe that were the basis for Garnavich et al. (1998a \cite{g1998a}), was augmented with SNe from the low redshift sample
gathered earlier at the Harvard-CFA (Riess 1996 \cite{riess1996}; Riess et al. 1999 \cite{retal1999}),
and by some from the Cal\'an--Tololo Survey (Hamuy et al. 1966 \cite{hetal1996}).
Between 27 and 34 SNe from these samples were used, depending on the criteria defined to constrain the quality of the light curves.
This small high redshift SN set was enough to clearly detect the Cosmic Acceleration (Riess et al. 1998 \cite{retal1998}).
The result was soon confirmed by the long awaited SCP full sample paper in 1999
(Perlmutter et al. 1999 \cite{p1999}), and by many subsequent papers from both teams.

\begin{figure}
\centering
\includegraphics[height=10cm]{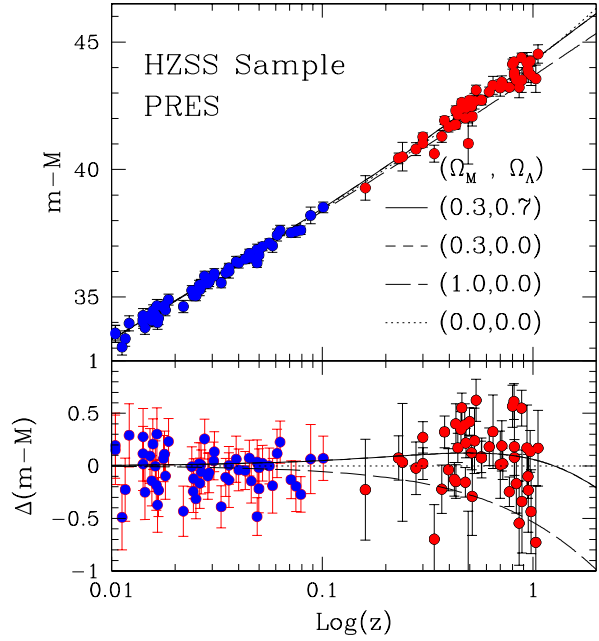}\vspace*{10mm}
%
%
\caption{Hubble Diagram of the 45 distant SNe collected by the High Z Supernova Search Team between 1995 and 2002 (red points), together with
58 SNe from the local sample (blue points). Distances for this plot were computed according with the PRES method (Prieto, Rest and Suntzeff 2003 \cite{prs2006}), an elegant generalization of the $\Delta m_{15}$ method.
The theoretical luminosity distances for universes with different density parameters $\Omega_M$, $\Omega_\Lambda$, are also shown, as in Figure
\protect{\ref{fi:D_L}}.
} \label{fi:HZTeam_sample}
\end{figure}

\begin{figure}
\centering
\includegraphics[height=10cm]{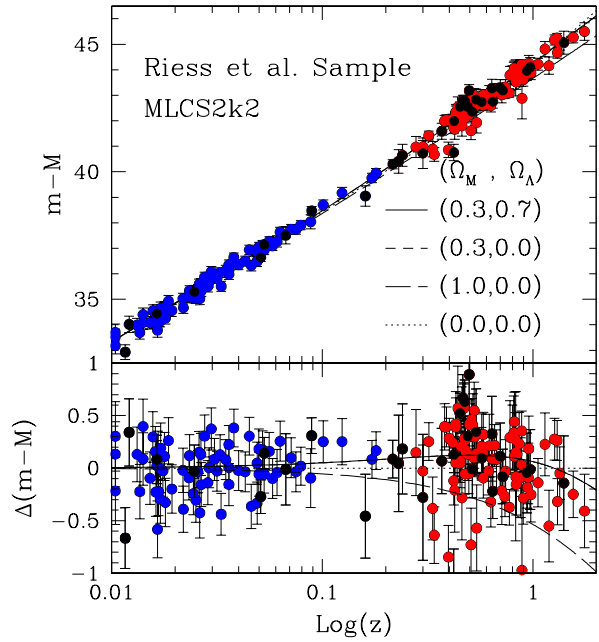}\vspace*{10mm}
%
%
\caption{Same as Figure~\protect{\ref{fi:HZTeam_sample}}, but for the 78 nearby and 109 distant SNe of the sample of Riess at al. 2004 \cite{retal2004} In this
case the distance were computed by the method MLCS2k2 (Jha et al. 2007 \cite{j2007}). 
Black symbols indicate SNe which do not qualify as "gold" according with the criteria of Riess et al. 2004 \cite{retal2004}.
} \label{fi:Riess_sample}
\end{figure}

What I will show here are the inferences based on the combined sample of distant SNe that the HZ Team collected throughout its existence
(presented in the previously mentioned papers and in Tonry et al. 2003 \cite{tetal2003}, Barris et al. 2004 \cite{betal2004}, and
Clocchiatti et al. 2006 \cite{cetal2006}),
and an expanded sample, including very distant SNe, built by Riess et al. 2004 \cite{retal2004}.
Figures \ref{fi:HZTeam_sample} and \ref{fi:Riess_sample} show, 75 nearby and 45 distant Type Ia SNe, and 78 nearby and 109 distant SNe, respectively. 
The SNe are color coded according to distance and quality of the light curves defined as in Riess et al (\cite{retal2004}.
The observed distance moduli are compared with the theoretical predictions for the same models of the universe shown in Figure \ref{fi:D_L}.

The simplest way to interpret the observed SN luminosity distances in empirical terms is to note that, at a redshift of $z \sim 0.5$ the points
tend to be above the theoretical line corresponding to the empty universe.
The distance moduli $m-M$ observed, in average, tend to be larger than those predicted,
even for a universe that will never decelerate due to its self gravitational attraction. 
So, SN are further away than expected according with the redshift of their parent galaxies. This observed {\em excess distance}
is interpreted by the models as the effect of an acceleration that has pushed the SNe farther
away between the time of their explosions at $z \sim 0.5$ and the present time at $z = 0$.

Also, it is important to note that, even with the spread of the points, the tendency of the SNe is to appear scattered around the empty universe at low
redshift (blue points), then preferentially above the empty universe (red points at $z \sim 0.45$ and then preferentially below the empty universe
at redshifts $z \sim 1$.
This behavior reinforces the strong signal of acceleration, and nullifies explanations based on systematic effects that will make distant
SNe to appear progressively more distant. This applies to the ``gray'' dust hypothesis (Aguirre 1999a \cite{a1999a}, 1999b \cite{a1999b}), but also
to most systematic effects, since they will typically show a monotonous trend with redshift.

\begin{figure}
\centering
\includegraphics[height=10cm]{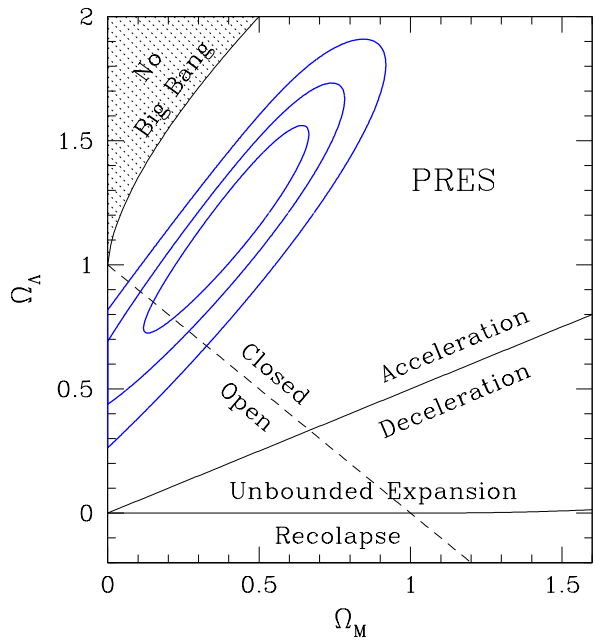}\vspace*{10mm}
%
%
\caption{Joint probability contours for the parameters $\Omega_{\Lambda}$ and $\Omega_M$ that best fit the Hubble diagram of
Figure~\protect{\ref{fi:HZTeam_sample}} (distances calibrated using the PRES method).
From larger to smaller, the drawn contours correspond to 99.5\%, 97\%, and 68\% confidence, respectively.
} \label{fi:HZSS_PRES_conts}
\end{figure}

\begin{figure}
\centering
\includegraphics[height=10cm]{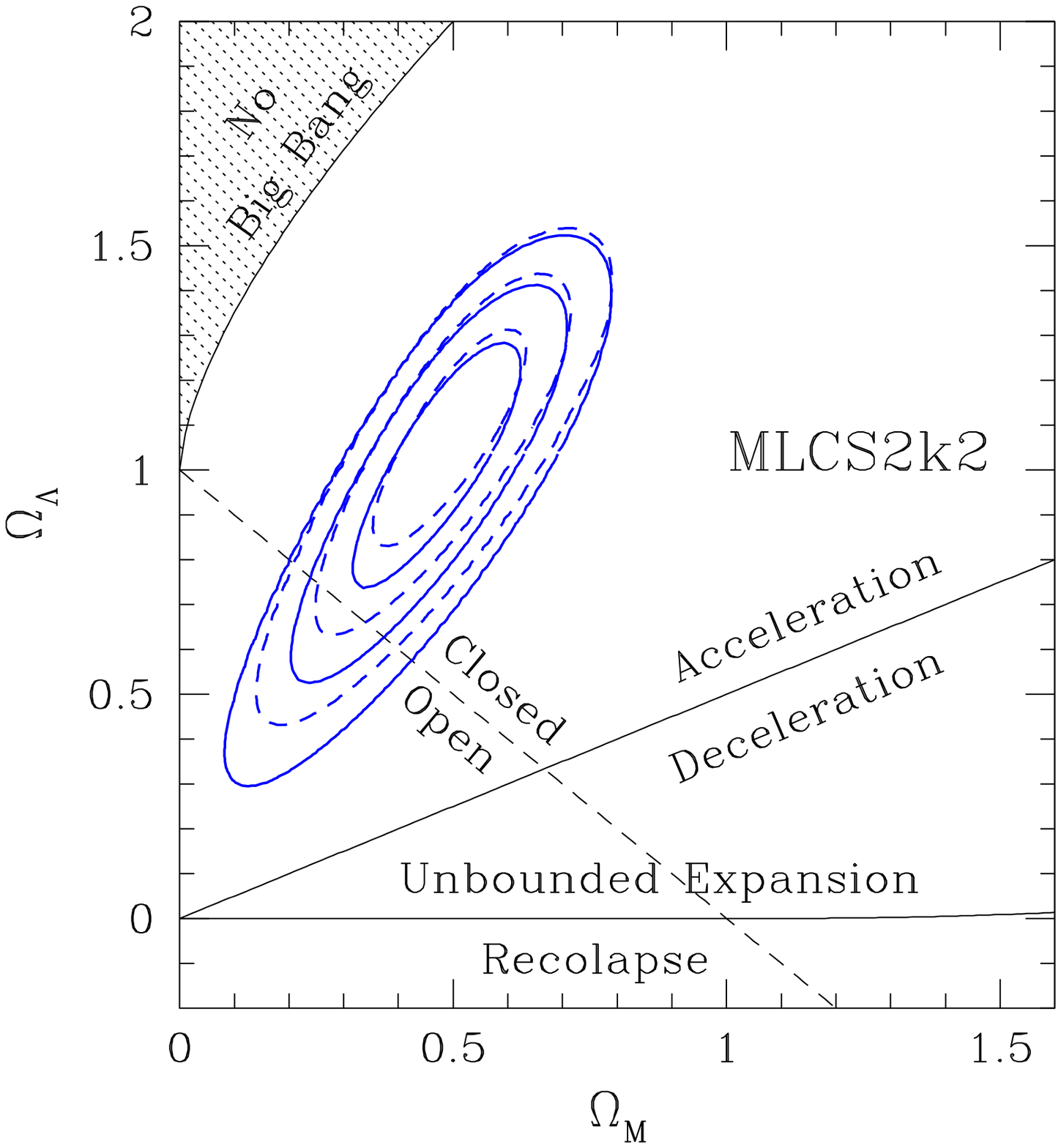}\vspace*{10mm}
%
%
\caption{Joint probability contours for the parameters $\Omega_{\Lambda}$ and $\Omega_M$ that best fit the Hubble diagram of
Figure~\protect{\ref{fi:Riess_sample}} (distances calibrated using the MLCS2k2 method).
The contours drawn with solid line include only those SNe that qualify as "gold" according with Riess et al. (2004 \cite{retal2004}).
The dashed contours include "gold" and "silver" SNe.
Again, from larger to smaller, the contours drawn correspond to 99.5\%, 97\%, and 68\% confidence, respectively.
} \label{fi:Full_MLCS2k2_conts}
\end{figure}

\subsection{The best fitting cosmology}

In the spirit of Gobbar et al. (1995 \cite{g1995}) it has become usual to asses the significance of the data for cosmology by using the Hubble Diagrams
to compute probability contours for arbitrary sets of cosmological parameters.
The particular process I will use is described in Riess et al. (1998). Basically, a set of $\Omega$ parameters, and an $H_0$ are assumed and a,
``observed minus expected from theory'' $\chi^2$ value is computed using equation \ref{eq:D_L}.
The probability of the resulting $\chi^2$ is obtained from the $\chi^2$ distribution and hipper-volume of $\chi^2$ density computed.
The volume is then dimensionally reduced by integrating over the ``nuisance'' parameters, of which $H_0$ is an example in this case.
Eventually, the volume can be projected in two dimensions like in the contour plots shown in Figures \ref{fi:HZSS_PRES_conts}
and \ref{fi:Full_MLCS2k2_conts}.

At first sight, either contour plot shows that the Hubble Diagrams alone do not impose a severe constraints on $\Omega_\Lambda$ or $\Omega_M$
individually, but they constrain the difference  $\Omega_M - \Omega_\Lambda$ rather tightly.
Even though, either contour level indicates with very high confidence that the Universe has $\Omega_\Lambda > 0$.
The ``excess distance'' described in graphical terms by the Hubble Diagrams is quantitatively interpreted, in terms of simple General Relativity,
by fitting a significant $\Omega_\Lambda$.

\begin{figure}
\centering
\includegraphics[height=10cm]{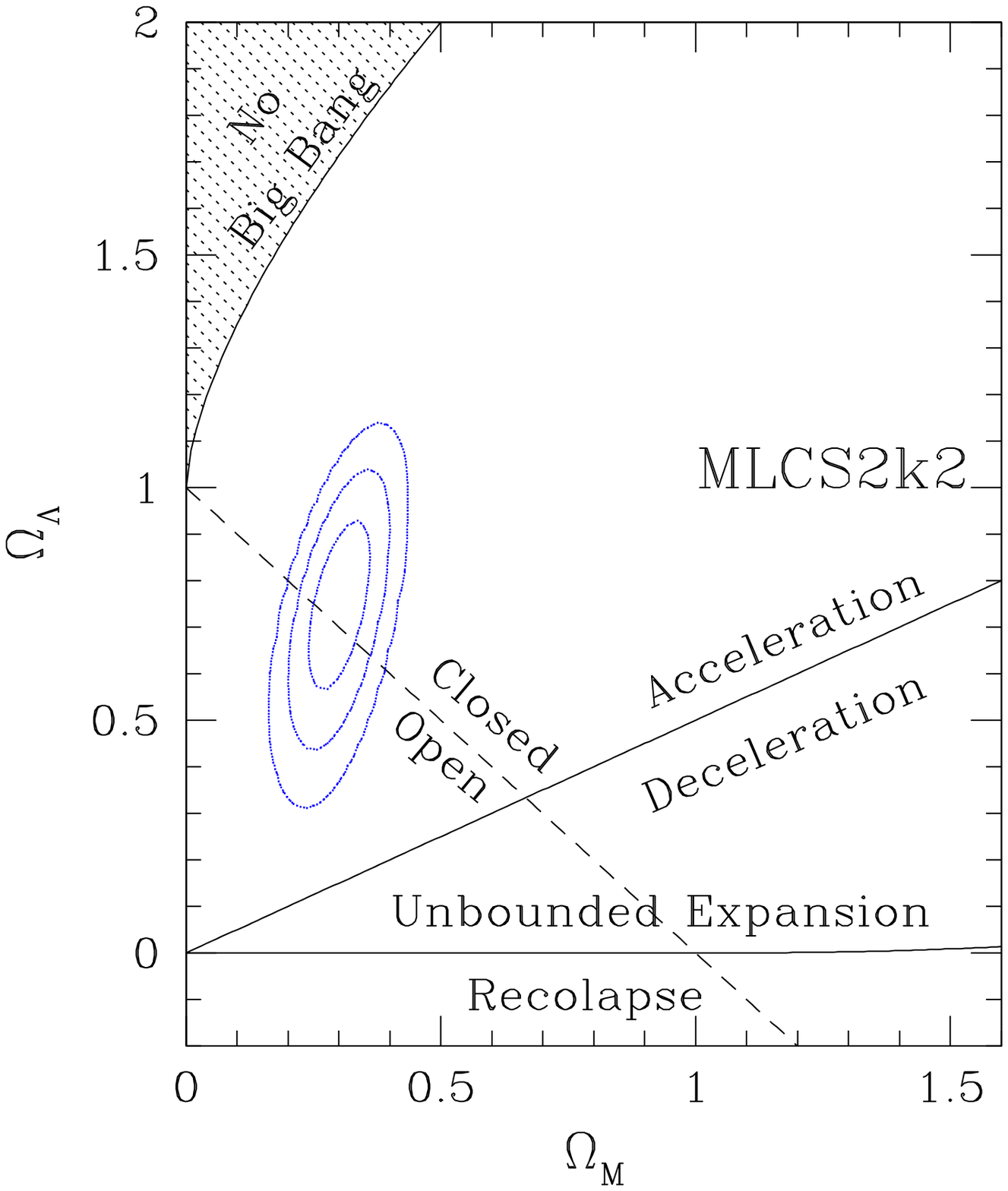}\vspace*{10mm}
%
%
\caption{Joint probability contours for the parameters $\Omega_{\Lambda}$ and $\Omega_M$ of the ``gold'' sample of 
Riess et al. (2004 \cite{retal2004}) (see previous figure), but now  computed assuming the 2dF prior in $\Omega_M$.
As before, from larger to smaller, the contours drawn correspond to 99.5\%, 97\%, and 68\% confidence, respectively.
} \label{fi:Full_MLCS2k2_ommprior_conts}
\end{figure}

The observational result is more compelling if the confidence contours from the Hubble Diagrams are combined with those of an independent experiment
to constrain either $\Omega_\Lambda$ or $\Omega_M$. In Figure \ref{fi:Full_MLCS2k2_ommprior_conts},
I show the outcome of this exercise assuming as a prior
$\Omega_M H_0/100 = 0.20 \pm 0.02$ as found by the Two Degree Field (2dF) Redshift Survey (Percival et al. 2001 \cite{petal2001}).
Just this two experiments point strongly to the parameters $\Omega_M \sim 0.3 $ and $ \Omega_\Lambda \sim 0.7$ which, together with $\Omega_K = 0$ and
your favorite prescription for the baryonic matter density, have been lately known as the ``Concordance Universe.''

\section{Is it $\Omega_\Lambda$ or a more general {\em Dark Energy}?}

\subsection{The ``Second Generation'' Surveys}

The success at detecting and measuring the Cosmological Constant using high precision Hubble Diagrams of distant Type Ia SNe prompted
astrophysicists to go a step further.
SNe appear now to be so precise as distance indicators that the goal of testing whether $\Omega_\Lambda$ is constant, or not, in time appears
to be possible.
On the theoretical side, many pressing questions are risen by a constant vacuum energy.
The more compelling appear to be the problems of scale and timing.
Carroll (2001 \cite{carroll2001}) presents a detailed review.
One of the possible solutions to the problems posed by a Cosmological Constant is that the ``Dark Energy'' is not just
a constant, but some other kind of field.

From the point of view of an observational astrophysicist, the experiment starts by
parametrizing the luminosity distances
in terms of this unknown energy density. It is natural to write the relation between pressure and density using the fairly
general form of the equation of state that applies to fluids
\begin{equation}
P = w \rho ,
\end{equation}
where $P$ is the pressure, $\rho$ the density, and $w$ a constant. From this form and the energy-momentum equation it is
found that $\rho$ evolves with the scale factor $a$ of the Universe as
\begin{equation}
\rho \propto a^{-n},
\end{equation}
where
\begin{equation}
n = 3 (1+w) .
\end{equation}
In this parametrization, gravitational matter ($n = 3$) requires $w = 0$, radiation ($n = 4$) implies $w = 1/3$, and
a Cosmological Constant ($n = 0$), will mean $w = -1$.

Introducing as before the cosmological density parameters by dividing the density of the different components into the
critical density, the equation \ref{eq:D_L} can be expressed as
\begin{equation} \label{eq:D_Lw}
D_L = \frac{c(1+z)}{H_0 \sqrt{|\Omega_k|}} \, \, {\rm sinn}
\left\{ \sqrt{|\Omega_k|} \int_0^z
\left[ \sum_i \Omega_i (1 + z')^{n_i} + \Omega_k (1 + z')^2
\right]^{-1/2} dz'
\right\},
\end{equation}
where the sub index $i$ denotes the different density components and the sub index $k$, as before, is reserved for
the curvature.

\begin{figure}
\centering
\includegraphics[height=10cm]{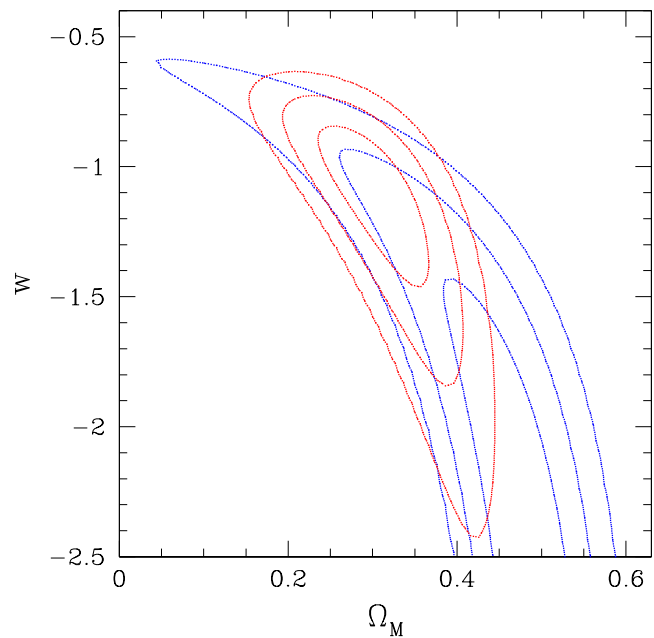}\vspace*{10mm}
%
%
\caption{Joint probability contours for the parameters $w$ and $\Omega_M$ for the ``gold'' sample of
Riess et al. (2004). The blue contours which extend below $w = -2.5$ are computed only with the SN sample. The red ones
were computed assuming the 2dF prior in $\Omega_M$.
As in all cases before, from larger to smaller, the contours drawn correspond to 99.5\%, 97\%, and 68\% confidence, respectively.
} \label{fi:UIA_contw_gold.eps}
\end{figure}

Equation \ref{eq:D_Lw} can be used in the same way as equation \ref{eq:D_L} to set up a minimization problem and simultaneously
constrain $\Omega_M$ and $w$, instead of $\Omega_M$ and $\Omega_\Lambda$.
I will use the sample of SNe presented before to illustrate the nature of the problem. Figure~\ref{fi:UIA_contw_gold.eps}
show the confidence contours computed with the ``gold'' sample of Riess et al. (2004 \cite{retal2004}) with and without the
additional constrain of an independent measurement of $\Omega_M$.
A quick inspection of the figure reveals that measuring $w$ is hard.
When applied to the problem of $\Omega_\Lambda$, the sample produces the results of Figures~\ref{fi:Full_MLCS2k2_conts} and
\ref{fi:Full_MLCS2k2_ommprior_conts}, which appear as a strong signal for the Cosmological Constant with $\Omega_\Lambda \sim 0.7$.
It is tough, however, to reach any meaningful conclusion about $w$ from the
contours of Figure~\ref{fi:UIA_contw_gold.eps}: the one sigma uncertainty is $\sim$60\%.
This kind of constrain is consistent with a very wide range of Dark Energy models.

The brute force approach to improve upon this result is to increase the number of good quality SN in the sample, from
many tens to a few hundred.
By the early 21st Century, two new collaborations appeared, with ex-members of both the SCP and the High Z SN Search Team rearranged among them,
and organized new projects to try to
answer the more complex question: Is the Cosmic Acceleration caused by a Cosmological Constant or not?
The newer projects were named SN Legacy Project (Astier et al. 2006 \cite{astier2006}) and ESSENCE (Miknaitis et al. 2007 \cite{miknaitis2007}).
Both collaborations have been working hard to understand the systematic biases, which, with the much increased number of SNe, will be the dominant source of uncertainty.
Interestingly, as the distant SN sample reaches the many hundreds to the thousands,
the nearby sample of around a hundred SNe becomes
one of the sources of systematic uncertainty.
Fortunately, too, some collaborations have also appeared to enlarge it (Hamuy et al. 2006 \cite{hetal2006}, Rau et al. 2009 \cite{retal2009}).

The early results, presented in several additional papers (Wood-Vasey et al. 2007 \cite{wood-vasey2007}, Guy et al.
2010 \cite{guy2010}, Sullivan et al. 2011 \cite{sullivan2011}) have taken the one sigma uncertainty down to $\sim$7\% and have been persintetly concluding
that $w$ is fully consistent with $-1$, this is, the plain Cosmological Constant.

\subsection{Conclusion}

Shall we call the problem solved and convince ourselves that the Cosmic Acceleration is, really, caused by a Cosmological Constant?
My personal view is that the case for it is strong, but still not certain.
As shown by the analysis of Miknaitis et al. (2007 \cite{miknaitis2007}),
we have not controlled yet all the possible systematic uncertainties
down to the very demanding limits required by this experiment.

A few years ago we confronted a similar situation when pondering just the existence of $\Omega_\Lambda$.
The theoretical expectations for it was so enormous that the very low observational
limits existing before 1998 were taken as a strong indication that $\Omega_\Lambda = 0$.
It was much easier to contrive mechanisms that will completely cancel it,
than imagine ways to make it much smaller than expected, but not zero.
Many of us were {\em convinced} that $\Omega_\Lambda = 0$.
Now, the observations indicate that $w \simeq -1$, making it difficult for us to resist the urge to jump at the conclusion that, then, $w = -1$.
We need to pay attention to the fact that the cosmic acceleration has fooled us more than
once in the past, and that there is still observational room for her to do it again.
The task for the observers during the next five to ten years is to develop better
instruments, better experiments, and to keep a paranoid eye on the sources of systematic
effects.

\section{Acknowledgements}

All of my work in the extremely exciting, demanding, challenging and competitive realm of
SN Cosmology has been done as a collaborator of two fairly large teams.
It started in 1995, when I was a postdoctoral fellow at Cerro Tololo Inter-American
Observatory in Chile.  I was invited by Brian Schmidt to join the High Z Supernova
Search Team, a now famous group of about 20 astrophysicists who were trying to measure
$q_0$, the {\em deceleration} parameter of the Universe.
It continued with the ESSENCE project, carried on  by an even larger team that grew
out of the HZ Team in 2001 and keeps on working to improve the SN constrains on cosmology.

Support for my research has been mainly provided by Chile, through many different grants. The active ones are
Iniciativa Cient\'{\i}fica Milenio P10-064-F (MINECON), Programa de Fondos Basales CATA PFB 06/09 (CONICYT), and
FONDAP No. 15010003 (CONICYT).




\bibliographystyle{aipproc}   


\end{document}